\documentclass[12pt,preprint]{aastex}

\begin{document}


\title{Diffractive Interaction and Scaling Violation 
in $pp \rightarrow \pi^0$ Interaction and GeV Excess in 
Galactic Diffuse Gamma-Ray Spectrum of EGRET}

\author{Tuneyoshi Kamae\altaffilmark{1},
Toshinori Abe
and Tatsumi Koi} 
\affil{Stanford Linear Accelerator Center, Menlo Park, CA 94025}
\email{kamae@slac.stanford.edu}

\altaffiltext{1}{Also with Kavli Institute for Particle Astrophysics
and Cosmology, Stanford University, Menlo Park, CA 94025}

\begin{abstract}
We present here a new calculation of the gamma-ray spectrum 
from $pp \rightarrow \pi^0$ in the Galactic ridge environment. 
The calculation includes the diffractive $p$-$p$ interaction
and incorporates the Feynman scaling violation for the first time.  
Galactic diffuse gamma-rays come, predominantly, 
from $\pi^0 \rightarrow \gamma \gamma$ in the sub-GeV to multi-GeV range.  
Hunter et al. found, however, an excess in the GeV range
(``GeV Excess'') in the EGRET Galactic diffuse spectrum 
above the prediction based on
experimental $pp \rightarrow \pi^0$ cross-sections and
the Feynman scaling hypothesis.
We show, in this work, that the diffractive process 
makes the gamma-ray spectrum harder than the incident proton
spectrum by $\sim 0.05$ in power-law index,
and, that the scaling violation produces 30$-$80\% more $\pi^0$ 
than the scaling model for incident proton
energies above 100~GeV.  Combination of the two can explain about a half 
of the ``GeV Excess'' with the local cosmic proton (power-law index
$\sim 2.7$).  The excess can be fully explained if the proton spectral 
index in the Galactic ridge is a little harder ($\sim 0.2$ in power-law
index) than the local spectrum.  Given also in the paper is that
the diffractive process enhances 
$e^+$ over $e^-$ and the scaling violation gives $50-100$\% 
higher $\bar{p}$ yield than without the violation, 
both in the multi-GeV range.

\end{abstract}



\setcounter{footnote}{1}
\section{Introduction}

Gamma-rays from neutral pions produced by 
cosmic-ray proton interactions with ISM have been 
predicted to dominate the diffuse Galactic emission 
in the sub-GeV to GeV band since 1960's. 
Early pioneers including \citet{Ginzburg67} and \citet{Hayakawa69} 
have estimated the gamma-ray flux
from $\pi^0$ together with other important mechanisms around that time.

First quantitative observation-based studies of 
the diffuse gamma-ray 
spectrum covering the sub-GeV and GeV band
were made, eg. by \citet{Strong78}, \citet{SB81}, \citet{Dermer86},
and \citet{Stecker89}.
They compared the data from COS-B 
\citep{COS-B} with their models based on experimental 
$pp \rightarrow \pi^0$ data from accelerators, their extension
to higher energies on the Feynman scaling hypothesis \citep{Feynman69},
estimations on cosmic ray proton and electron fluxes, 
and the ISM distribution obtained by radio surveys. 
Within the uncertainties in the data and modeling, 
the studies cited above concluded that gamma-rays from $\pi^0$ 
are a dominant component in the Galactic ridge spectrum above 100 MeV.
The bremsstrahlung emission by $e^+/e^-$ off 
ISM atoms and the inverse-Compton scattering of infra-red and
optical photons by $e^+/e^-$ are also expected to contribute 
significantly in the sub-GeV to GeV
energy range \citep{Hayakawa69,MurthyWolfendale93,Schoenfelder01} 

The limited statistics and 
energy coverage of the COS-B gamma-ray data permitted
only a crude consistency check of the
$pp \rightarrow \pi^0 \rightarrow \gamma$ hypothesis
within a factor $\sim 2$ and
left large ambiguity on the mix of
emission mechanisms \citep{SB81}.  The spatial distribution of the 
energy-integrated gamma-ray intensity, on the other hand, 
gave a higher statistical accuracy on which 
\citet{Hasselwander82}, \citet{Strong82}, \citet{Bloemen84}, 
\citet{Bloemen85}, and others set the path to the 
Galactic gamma-ray astronomy.  We refer to  
\citet{MurthyWolfendale93}, \citet{Schoenfelder01}, and 
\citet{Schlickeiser02} for general references on the topics of this work.
  
When the much improved data obtained with EGRET \citep{EGRET} 
were studied by \citet{Bertsch93} and \citet{Hunter97}, an excess of about 
$\times (1.5-2)$ became apparent in the data in the GeV band relative to
the Galactic gamma-ray emission models cited above. 
This excess is visible along the entire Galactic plane 
($-10<b<10$ deg. and $-90<\ell<+90$ deg.) 
but most pronounced in the central region of the plane
($-40<\ell<+40$ deg.).  Here $b$ and $\ell$ 
are the Galactic latitude and longitude, respectively. 
In literature, the excess is referred to as the ``GeV Excess''.

\citet{Mori97} studied the diffuse gamma-ray emission by using
Monte Carlo $p$-$p$ interaction simulators devloped for accelerator
experiments and confirmed that the EGRET spectrum can not be reproduced
with the conventional cosmic proton spectrum.

Strong and Moskalenko developed a computer program, 
GALPROP, where the cosmic ray propagation and interaction are 
numerically calculated in a model Galaxy 
to obtain the spatial distribution and spectrum of 
secondary particles including gamma-rays 
and radio isotopes \citep{Galprop1,Galprop2}.  
The diffuse Galactic gamma-ray spectra have been calculated 
separately for the $\pi^0$ decay, bremsstrahlung, and inverse-Compton
with various parameter settings of GALPROP  
by \citet{SMR00}.  They noted that: a) the local interstellar spectra of 
electrons and protons (power-law index $\ge 2.5$ above $10-20$ GeV)
do not reproduce the gamma-ray spectrum along the Galactic plane
observed by EGRET; b) a very hard electron spectrum 
(power-law index $\sim 1.8$) and a modified nucleon spectrum will be 
needed to minimize the difference between the prediction and
the data in the GeV band; and
c) the GeV excess in the central Galactic ridge can not be reproduced
within the constraint on the proton spectral index imposed 
by recent cosmic proton measurements and the limit on the electron 
spectral index ($\sim 1.9$) from radio and local cosmic ray
observations.  They noted that the excess persists  
at higher Galactic latitude ($|\ell|>5$ deg).

\citet{Buesching01} have also noted that 
the EGRET spectrum is incompatible with the locally measured 
cosmic proton spectrum.  

The GeV Excess has led to new 
optimizations of the Galactic gamma-ray emission models 
and speculations on possible
new gamma-ray sources in the Galaxy.
One choice is to assume a harder proton spectrum
in the Galactic ridge region as has been noted by \citet{Mori97}.  
Another choice is to assume a much higher electron flux
with a broken power-law spectrum in the Galactic ridge \citep{SMR04}.  
\citet{Buesching01} has proposed to introduce a mix of spectral indices 
for protons which leads to a convex gamma-ray
spectrum similar to that observed by EGRET.
Possible contributions of unidentified pulsars \citep{Pohl97}
and dark-matter particle annihilation \citep{deBoer03,
Cesarini04} have also been studied. 

We came to note that all calculations of 
the Galactic gamma-ray emission cited above have not included 
an important component of inelastic $p$-$p$ interaction, 
the diffractive interaction, nor incorporated 
the Feynman scaling violation in the non-diffractive inelastic interaction.
Another important finding was that these calculations assume an obsolete
$p$-$p$ non-diffractive inelastic cross-section model taking
a constant value of $\sim 24$~mb for $T_p \gg 10$~GeV.
Updating these shortfalls and inaccuracy can change the 
gamma-ray spectrum from the proton ISM interaction at high energies in the
following ways: the diffractive process will add
gamma-rays in the highest end of the spectrum;
the scaling violation and the up-to-date non-diffractive 
inelastic cross-section will increase the
gamma-ray yield in the GeV to multi-GeV range.


We built a model (model A) representing the latest knowledge on the $p$-$p$
inelastic interaction, and calculated the
gamma-ray spectrum due to $\pi^0$ produced in the
cosmic-ray proton interaction with ISM.  
The $p$-$p$ interaction is simulated 
separately for the non-diffractive inelastic  
and for the diffractive processes.  
The non-diffractive process is calculated  
by two computer programs: Pythia 6.2 by \citet{Pythia62} 
for the proton kinetic energy ($T_p$) range 
$512$~TeV $\ge T_{p} \ge 52.6$~GeV and 
the model by \citet{SB81}
with the parametrization of \citet{Blattnig00} 
for $52.6$~GeV $\ge T_{p} > 0.488$~GeV.
The diffractive process is simulated by a program 
written for this study \footnote{Simulation Program 
DiffDissocSimulNew.py for model A is available upon request
from author.}: 
it is based on the formulae 
given in \citet{Goulianos83}, \citet{Goulianos95}, and \citet{Goulianos99}.
The non-diffractive and diffractive gamma-ray spectra are added
according to the cross-section model for model A shown in 
Columns 2 and 3 of Table 1 and Fig.1a\footnote{The first 
2 data sets in Figure 1 are from \citet{Hagiwara02} and 
the last from the Spires database (see http://www.slac.stanford.edu/spires).}.

\begin{small}

\begin{table}
\caption{$p$-$p$ Model Cross-Sections and Galactic Proton Spectral Models}
\begin{tabular}{rrrrrrr}
\hline\hline
$T_p$ & \multicolumn{2}{c}{$\sigma$(model A)} &
           $\sigma$(model B) &
             \multicolumn{3}{c}{Factors for Proton Spectra} \\
\hline
(GeV) & $\sigma$(NonDiff) & $\sigma$(Diff) & $\sigma$(NonDiff) 
              & Ind=2.0 & LIS & Trial4GR \\
\hline
4.88E-01 & 5 & 0 & 5 & 2.05E+03 & 5.54E+04 & 3.05E+04\\
6.90E-01 & 20 & 0 & 20 & 1.45E+03 & 4.58E+04 & 2.01E+04\\
9.80E-01 & 23.4 & 0 & 23.4 & 1.02E+03 & 3.72E+04 & 1.32E+04\\
1.38E+00 & 25 & 0 & 24 & 7.25E+02 & 2.91E+04 & 8.75E+03\\
1.95E+00 & 27.57 & 0 & 24.6 & 5.13E+02 & 2.13E+04 & 5.78E+03\\
2.76E+00 & 28.57 & 0 & 24.4 & 3.62E+02 & 1.49E+04 & 3.81E+03\\
3.91E+00 & 29.27 & 0 & 24.1 & 2.56E+02 & 9.54E+03 & 2.51E+03\\
5.52E+00 & 29.76 & 0 & 23.8 & 1.81E+02 & 5.85E+03 & 1.66E+03\\
7.81E+00 & 23.96 & 6.13 & 23.4 & 1.28E+02 & 3.51E+03 & 1.09E+03\\
1.11E+01 & 23.65 & 6.68 & 23 & 9.05E+01 & 2.04E+03 & 7.21E+02\\
1.56E+01 & 23.29 & 7.22 & 22.6 & 6.40E+01 & 1.18E+03 & 4.75E+02\\
2.21E+01 & 22.91 & 7.76 & 22 & 4.53E+01 & 6.52E+02 & 3.04E+02\\
3.13E+01 & 22.53 & 8.29 & 21.6 & 3.20E+01 & 3.62E+02 & 1.81E+02\\
4.42E+01 & 22.27 & 8.72 & 21.6 & 2.26E+01 & 2.01E+02 & 1.08E+02\\
6.25E+01 & 22.23 & 8.96 & 21.6 & 1.60E+01 & 1.11E+02 & 6.40E+01\\
8.84E+01 & 22.2 & 9.19 & 21.6 & 1.13E+01 & 6.18E+01 & 3.81E+01\\
1.25E+02 & 22.14 & 9.43 & 21.6 & 8.00E+00 & 3.43E+01 & 2.26E+01\\
1.77E+02 & 22.14 & 9.67 & 21.6 & 5.66E+00 & 1.90E+01 & 1.35E+01\\
2.50E+02 & 22.22 & 9.91 & 21.6 & 4.00E+00 & 1.06E+01 & 8.00E+00\\
3.54E+02 & 22.36 & 10.15 & 21.6 & 2.83E+00 & 5.86E+00 & 4.76E+00\\
5.00E+02 & 22.58 & 10.39 & 21.6 & 2.00E+00 & 3.25E+00 & 2.83E+00\\
7.07E+02 & 22.88 & 10.64 & 21.6 & 1.41E+00 & 1.80E+00 & 1.68E+00\\
\hline
1.00E+03 & 23.24 & 10.88 & 21.6 & 1.00E+00 & 1.00E+00 & 1.00E+00\\
\hline
1.41E+03 & 23.67 & 11.12 & 21.6 & 7.09E-01 & 5.58E-01 & 5.97E-01\\
2.00E+03 & 24.18 & 11.36 & 21.6 & 5.00E-01 & 3.08E-01 & 3.54E-01\\
2.80E+03 & 24.75 & 11.6 & 21.6 & 3.57E-01 & 1.74E-01 & 2.13E-01\\
4.00E+03 & 25.4 & 11.85 & 21.6 & 2.50E-01 & 9.47E-02 & 1.25E-01\\
5.66E+03 & 26.1 & 12.09 & 21.6 & 1.77E-01 & 5.25E-02 & 7.43E-02\\
8.00E+03 & 26.88 & 12.33 & 21.6 & 1.25E-01 & 2.92E-02 & 4.42E-02\\
1.13E+04 & 27.72 & 12.57 & 21.6 & 8.85E-02 & 1.62E-02 & 2.63E-02\\
1.60E+04 & 28.63 & 12.82 & 21.6 & 6.25E-02 & 8.97E-03 & 1.56E-02\\
2.26E+04 & 29.6 & 13.06 & 21.6 & 4.42E-02 & 4.98E-03 & 9.29E-03\\
3.20E+04 & 30.64 & 13.3 & 21.6 & 3.13E-02 & 2.76E-03 & 5.52E-03\\
4.53E+04 & 31.74 & 13.54 & 21.6 & 2.21E-02 & 1.53E-03 & 3.29E-03\\
6.40E+04 & 32.9 & 13.79 & 21.6 & 1.56E-02 & 8.50E-04 & 1.95E-03\\
9.05E+04 & 34.12 & 14.03 & 21.6 & 1.11E-02 & 4.72E-04 & 1.16E-03\\
1.28E+05 & 35.41 & 14.27 & 21.6 & 7.81E-03 & 2.62E-04 & 6.91E-04\\
1.81E+05 & 36.76 & 14.51 & 21.6 & 5.53E-03 & 1.45E-04 & 4.11E-04\\
2.56E+05 & 38.18 & 14.76 & 21.6 & 3.91E-03 & 8.05E-05 & 2.44E-04\\
3.62E+05 & 39.65 & 15 & 21.6 & 2.76E-03 & 4.47E-05 & 1.45E-04\\
5.12E+05 & 41.19 & 15.24 & 21.6 & 1.95E-03 & 2.48E-05 & 8.63E-05\\
\hline\hline
\end{tabular}
\end{table}
\end{small}

\begin{figure}
\epsscale{.80}
\plotone{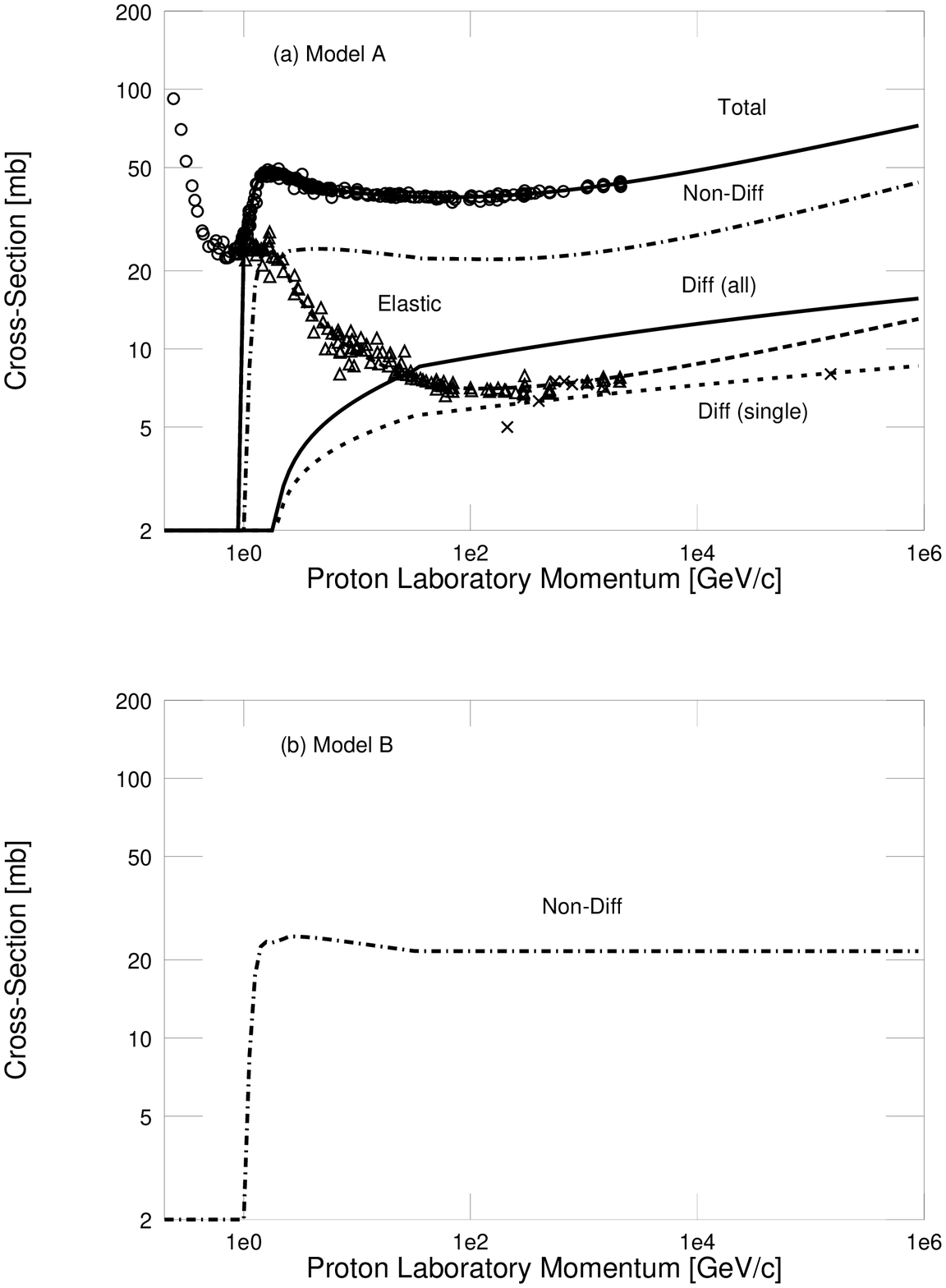}
\caption{The $p$-$p$ cross-section models: (a) for model A and (b) 
for model B. Curves are for the total ({\it{upper solid}}), non-diffractive 
inelastic ({\it{dot-dashed}}), elastic ({\it{dashed}}), 
all diffractive ({\it{lower solid}}), 
and single diffractive ({\it{dotted}}) processes.  
Note that model B is made only of
non-diffractive inelastic process.
Data are for the total ({\it{circles}}),
elastic ({\it{triangles}}), and single diffraction ({\it{crosses}}).
\label{fig1}}
\end{figure}

The gamma-ray spectrum from $pp \rightarrow \pi^0$
will be presented for 3 cosmic proton spectra: power-law spectrum of index 2.0
representing the acceleration site (see, eg., \citet{Ellison04}); 
the local interstellar spectrum 
(referred to as LIS) obtained from the recent primary 
cosmic proton measurements (see eg. \citet{MSOP02}); 
and a trial broken power-law
spectrum for the Galactic ridge region
(index=2.5 and 2.2 for above and below $T_p=20$ GeV, respectively).

We also built a reference model, model B, 
consisting only of the non-diffractive process 
as all previous models have assumed \citep{Strong78,SB81,Dermer86,Stecker89}. 
The non-diffractive cross-section assumed in model B approaches 
a constant value (see Column 4 of Table 1 and Fig.1b) 
similar to that assumed in the above references.
The computer programs for the model A non-diffractive process 
are used to calculate the model B non-diffractive process
except for the Pythia parameters as will be described later in Sec.5.

In the following sections, we will describe how 
the total inelastic cross-section is broken down in our model A;
relate the Feynman scaling hypothesis and numerical simulation codes 
in a historic perspective; describe how the
scaling violation is incorporated in Pythia; introduce our simulation 
code of the diffractive interaction; and 
present the gamma-ray spectra obtained with 
models A and B.
We then compare the predictions of model A, \citet{SB81},
\citet{SMR04} and model B 
with the EGRET gamma-ray spectrum (EGRET Archive)
\footnote{See EGRET archived data at 
ftp://cossc.gsfc.nasa.gov/compton/data/egret/high\_level/combined\_data/\\
galactic/counts.g1234\_30.g001.
We have subtracted all point sources listed in the 
EGRET 3rd Catalog from the data above to make  
the diffuse gamma-ray spectrum in the Galactic ridge.
This point-source subtracted data will be made available 
in the publication in preparation.}

Finally discussion on the results, our conclusions, and 
possible implications on $\bar{p}$,
neutrino, $e^+$, and $e^-$ spectra produced by $p$-$p$ interactions
will be given.
Technical details on models A and B will be given in Appendix. 

\section{Breakdown of the Inelastic Cross-Section}

In the present work,
the total proton-proton cross-section is broken down to 
the elastic, non-diffractive inelastic, and 
diffractive inelastic cross-sections.  
The total and elastic cross-sections have been measured accurately
in the proton kinetic energy range relevant to this
study as compilated in \citet{Hagiwara02}\footnote{Data on the 
total and elastic cross-sections are available at
http://pdg.lbl.gov/2002/contents\_plots.html.}
They are plotted as experimental points in Fig.1a together 
with the cross-sections used in model A.  
The total inelastic cross-section is, 
by definition, the difference between the two cross-sections.  

Experiments at CERN-ISR established, in mid 1970's, that 
the total, elastic, and inelastic cross-sections increase with 
the incident proton energy \citep{CERN-ISR1,CERN-ISR2,CERN-ISR3}.  
A class of inelastic interaction where the projectile proton
and/or the target proton transition to excited states (discrete 
nucleon resonances and conituum) became known 
by early 1970's \citep{AlberiGoggi81}.  
This new class of interaction, the diffractive interaction,
has been found to grow in cross-section 
with the incident proton energy 
\citep{DiffDissoc1,DiffDissoc2,DiffDissoc3,DiffDissoc4,DiffDissoc5,
DiffDissoc6}.  When only one proton transitions to an
excited state, the process is called the single
diffractive interaction: otherwise the double diffractive interaction.  
The diffractive process will be 
explained later in Sec.5.

The early data led to a naive conjecture that the diffractive
cross-section increases with the incident proton energy while 
the non-diffractive inelastic cross section stays constant 
at $\sim 21-24$ mb above $\sim 10$~GeV.
According to recent studies, this conjecture is
over-simplification and inaccurate.  The increase in the 
total cross-section is shared by the non-diffractive and 
diffractive processes
as incorporated in the model A cross-sections and shown in 
Fig.1a \citep{Goulianos95,Goulianos99,CDF01}.
Fig.1b shows the model B non-diffractive inelastic cross-section:
here the diffractive process is completely neglected as has been 
in all previous predictions on the diffuse Galactic gamma-ray 
spectrum \citep{Strong78,SB81,Dermer86,Stecker89,Mori97}.     

\section{Scaling Hypothesis and Simulation of Non-Diffractive Interaction}

Scaling hypotheses, or scaling, have been introduced 
in many branches of physics in many different contexts.
Besides the Feynman scaling, there is one more well-known 
scaling hypothesis for the high energy particle
interaction the Bjorken scaling \citep{Bjorken69,TextBook1,TextBook2}.
The hypothesis by Bjorken refers to the lepton-initiated 
deep inelastic scattering on a nucleon while that by Feynman deals 
with hadronic interactions.  
We apply the latter to the $pp \rightarrow \pi^0 X$
cross-section, where $X$ stands for all states 
reachable from the initial state.  The cross-section of this kind 
is referred to as the ($\pi^0$) inclusive cross-section.
Historically the Feynman scaling was used
to supplement lack of data at higher energies 
\citep{Strong78,SB81,Dermer86,Stecker89,Naito94,Hunter97,
Mori97,SMR00,Buesching01,SMR04}.  At present, 
computer-based models are available,
such as Pythia \citep{Pythia62}, where 
experimental data up to $T_p=$ a few 100~TeV are reproduced.
In this section, we will summarize briefly 
how such a computer program (eg. Pythia) simulates 
high energy quark-parton interactions and subsequent hadronization.

The scaling hypothesis by Feynman \citep{Feynman69} applies to the 
non-diffractive interaction at high energies ($T_p \gg 10$ GeV). 
The hypothesis states that cross-sections depend only
on the scaling variable $x^*_{||} = 2 p^*_{||}/\sqrt{s}$ 
and $p_{T}$ at the high energy limit.   
Here $p^*_{||}$ is the momentum component parallel to 
the relative motion between the projectile and the target, 
$p_{T}$ the perpendicular component, and $\sqrt{s}$ the total energy 
in the center-of-mass system.  This hypothesis has become
a powerful tool in extrapolating low energy data to higher
energies inaccessible by accelerators \citep{TextBook1,TextBook2}.

Feynman proposed the parton model \citep{Feynman72}
as a physical model realizing the scaling hypothesis. 
The parton model soon became the quark-parton model
and enhanced its predictability.  
It is, however, the perturbative QCD 
by Altarelli and Parisi \citep{pQCD} that gave 
a broader foundation for calculating the absolute cross-section 
of complex final states.
On this basis, Anderson and collaborators started, in late 1970's,
the developmental work toward a widely used numerical
simulation code for $e^+e^-$ collider experiments, 
the Lund model \citep{LundModel1,LundModel2,LundModel3}. 
Another equally popular simulation code, Herwig, was written 
a few years later by Webber and Marchesini 
\citep{Herwig1,Herwig2,Herwig3} on a different algorithm known as 
the Jet Calculus \citep{JetCalculus1,JetCalculus2}.  
Pythia \citep{Pythia62} is considered as an extension of the Lund model
for high energy $p$-$p$ and $p$-$\bar{p}$ interactions. 
Herwig has also been extended to a similar direction \citep{Herwig65}.
We note that simulation codes have been developed for
hadron-hadron, hadron-nucleus, neucleus-neucleus, and photon-nucleus
interactions including PHOJET\footnote{See R. Engel 1997 
at http://www-ik.fzk.de/\~engel/phojet.html.}.
However they are not as widely used as Pythia and Herwig 
in simulating high energy $p$-$p$ interactions.
These simulation codes have evolved, cross-checking mutually 
as well as against experimental data, and built up confidence
in the computer-based simulation over 20 years
(see, eg., \citet{Sjostrand99}, \citet{MrennaRichardson03}
and references therein).

Since both Pythia and Herwig are written on the perturbative QCD, the
scaling is violated when a hard parton-parton interaction occurs.
However there are many more parton diagrams that
violate the Feynman scaling as will be discussed in Section 4.
A set of such higher order terms have been added to
Pythia recently \citep{Sjostrand04}. 
We use Pythia without such higher order terms 
as a reference model (model B) that will
approximate, crudely, the Feynman scaling model.  
We warn, however, that Pythia without the higher order terms 
(our model B) and the scaling models such as \citet{SB81} 
and \citet{Dermer86} give different gamma-ray spectral shapes 
as has been shown in Fig.3 of \citet{Mori97}.   

Table 2 tabulates the computer programs and the cross-section
models used in this work.

\begin{table}
\begin{center}
\caption{Models of $p$-$p$ interactions and Cross-Sections}
\end{center}

\begin{tabular}{llll}
\hline\hline
 & \multicolumn{2}{c}{Non-Diffractive} 
                         & Diffractive \\
Proton K.E. ($T_p$) & Model A & Model B & Model A \\
\hline\hline
\multicolumn{2}{l}{Cross-Section Models} & & \\ 
\hline
0.410GeV$-$609TeV & Table 1 col. 2 & Table 1 col.4
                         & Table 1 col. 3 \\
              & Fig.1a Non-Diff & Fig.1b Non-Diff 
                         & Fig.1a Diff(all) \\
\hline\hline
\multicolumn{2}{l}{Interaction Models} & & \\
\hline
0.410$-$52.6~GeV  &  \multicolumn{2}{c}{\citet{SB81}}
                  & DiffDissocSimNew.py\tablenotemark{a}  \\
                  &  \multicolumn{2}{c}{Eqs.23/24 and eq.32 of} 
                         & T. Kamae(2004, per. comm.)\tablenotemark{a} \\
                  &  \multicolumn{2}{c}{and \citet{Blattnig00}} & \\
\hline
52.6GeV$-$609TeV
              &  Pythia6.2 & Pythia6.1
                         & T. Kamae(2004, pers. comm.)\tablenotemark{a} \\
              &  (Higher order) & (No higher order) & \\
              &  (``Tune A'') &    & \\
\hline
\end{tabular}
\tablenotetext{a}{Simulation program DiffDissocSimulNew.py for model A 
is available upon request from author.}
\end{table}

\begin{figure}
\epsscale{.80}
\plotone{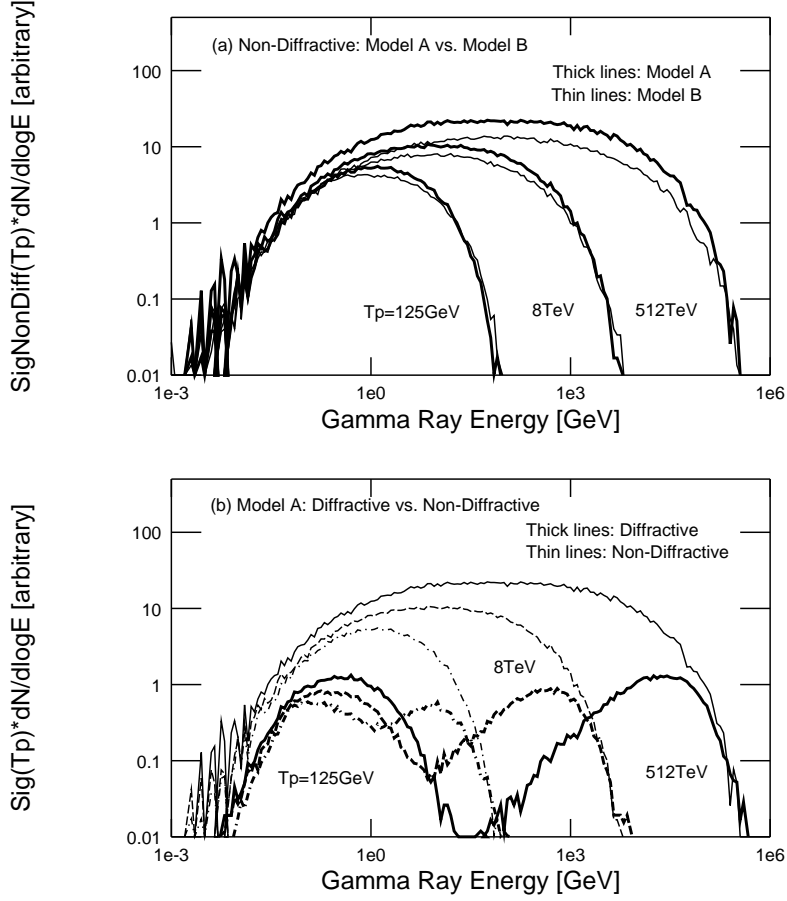}
\caption{Predicted gamma-ray spectra for 3 mono-energetic
proton beams: (a) the non-diffractive 
contribution in model A ({\it{thick lines}}) and model B ({\it{thin lines}});
and (b) the non-diffractive in model A ({\it{thick lines}}) 
and diffractive in model A ({\it{thin lines}}).  
Proton kinetic energies ($T_p$)
are 512~TeV (solid), 8~TeV ({\it{dashed}}), and 125~GeV ({\it{dot-dashed}}).
Note that model A generates $30-80$~\% more multi-GeV
gamma-rays for $T_p>100$~GeV. \label{fig2}}
\end{figure}

Several corollary scaling laws have been derived from 
the quark-parton model.  Of them, the best known and relevant 
to the present paper is the KNO scaling in the particle 
multiplicity distribution \citep{KNO72}.
According to the KNO scaling, the distribution of 
the number of secondary particles (eg. $\pi^0$ or $\pi^{+/-}$) 
takes a common shape when re-scaled by the averaged multiplicity
of the particle.
This scaling law has been used to detect and quantify
the scaling violation in the charged particle multiplicity
distribution.

\section{Scaling Violation in Non-Diffractive Interaction}

The two scaling hypotheses of particle physics,
the Feynman scaling \citep{Feynman69} and 
the Bjorken scaling \citep{Bjorken69}, are known to be
violated.  In literature they are simply referred to as
the scale violation.  The final 
states the two hypotheses deal with are both controlled by QCD,
but their violation originates from its two different aspects
\citep{TextBook1}.

The scaling violation in the $p$-$\bar{p}$ inclusive cross-section  
has become evident as the center-of-mass energy reached multi-TeV 
at CERN-SPS in 1984 \citep{KNOviolation83,KNOviolation84,
Breakstone84},
almost one decade after the violation of the Bjorken scaling had been
established.  The violation has manifested itself 
as deviation from the KNO scaling in the multiplicity distribution
and excess of jets with large transverse momentuma. 
The scaling is now believed to be violated because both 
the perturbative QCD and the multiple parton interaction 
introduce new degrees of freedom other than
the scaling variable $x$ \citep{Sjostrand87,Sjostrand04}.  
Such complex processes are best studied by computer simulations:
scaling violation due to the perturbative QCD has been 
included in Pythia from its birth but the multiple parton interaction terms
(will be referred to as the higher order terms) have been added
recently, eg. in Version 6.2 we use for model A \citep{Pythia62}.  

The CDF collaboration at Fermi Lab. has tuned the parameters 
controlling the multiple parton interaction in Pythia 6.2 and
made the parameter set available as Tune A
\citep{CDFtuneA,Sjostrand04}.  We use Pythia 6.2 with this
parameter set as the non-diffractive part of model A for 
$T_p\ge 62.5$~GeV (see Appendix 1 for the parameter setting).
We note that we use Pythia 6.1 without the higher 
order terms in model B: the gamma-ray spectra by the non-diffractive 
parts of models A and B are compared in Fig.2a.  
The $\pi^{+/-/0}$ inclusive cross-sections produced by
Pythia 6.1 and 6.2 have been verified to agree when they are ran 
without the higher order terms.
The difference in magnitude between models A and B 
non-diffractive contributions 
is mostly due to the difference in the two
non-diffractive cross-sections (see in Table 1 and Figs.1a and 1b).  
The scaling violation manifests itself as an increase 
in the gamma-ray yield and a subtle change in the spectral shape.  
We defer comparison between the non-diffractive and 
diffractive parts of model A shown in Fig.2b to Section 5.  
See Appendix for further details on the Pythia parameter choices.
   
The scaling violation affects the charged particle 
multiplicity distribution and violates the KNO scaling: 
the measured distribution at
$T_p =$21.3~TeV ($E_{CM} =$200~GeV) by \citet{Ansorge89} is compared 
in Fig.3, with 
model ``A All'', model A without diffractive interaction, and model B.
Since the experiment did not trigger on the single diffractive process,
we removed the single diffraction contribution in model ``A All''.  
The KNO scaling based on the data taken at $E_{CM}=30.4$ GeV
($T_p=490$ GeV) \citep{Breakstone84} predicts a skewed normal distribution 
with mean at 22.3 and mean deviation 11.7, which is close to  
the model B curve in Fig.3.  Important is to note that 
the model B curve also violates the KNO scaling when compared 
with the charge multiplicity 
distribution at $T_p<50$~GeV \citep{Dao73}.

The charged multiplicity distribution by \citet{Ansorge89}
shown in Fig.3 is that by the antiproton-proton interaction.  
In fact high energy ($T_p>1$~TeV)
experimental data to verify our modeling are only 
available from $p$-$\bar{p}$ collider experiments at CERN-SPS and 
Tevatron Collider at Fermi Laboratory.  This substitution
of $p$-$p$ by $p$-$\bar{p}$
is well-founded experimentally up to $E_{CM}=50$~GeV \citep{Hagiwara02}
and theoretically for $E_{CM}>50$~GeV (see eg. \citet{TextBook1}).
The difference between particle- and antiparticle-induced
cross-sections are predicted to diminish asymptotically 
\citep{TextBook1,Hagiwara02}.  We have confirmed that the
$pp \rightarrow \pi$ and $\bar{p}p \rightarrow \pi$ cross-sections
calculated by Pythia agree within statistical errors at $21.3$~TeV.  

\begin{figure}
\epsscale{.80}
\plotone{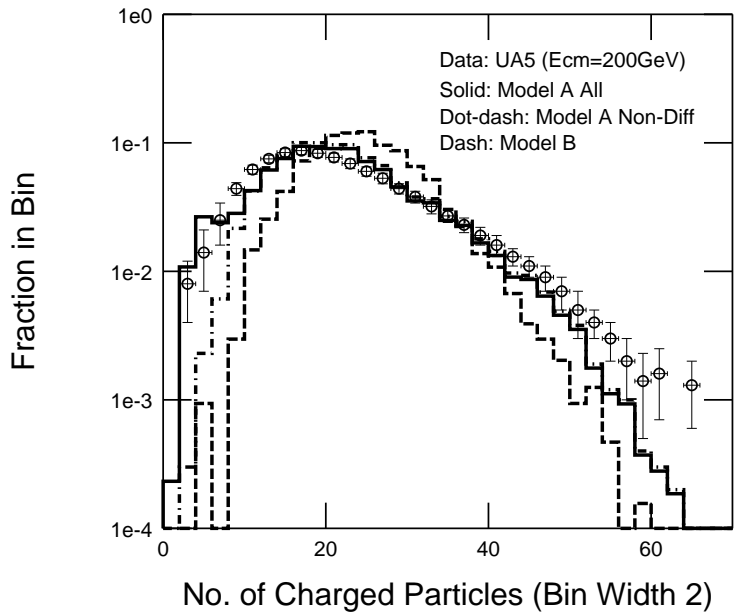}
\caption{Charged particle multiplicity distribution for $p$-$\bar{p}$ 
inelastic interaction at $E_{lab}=21.3$ TeV ($E_{CM}=200$ GeV).
Histograms are: ({\it{solid}}) model A non-diffractive + double diffractive, 
({\it{dot-dashed}}) model A non-diffractive, and ({\it{dashed}}) model B.  
Data are from \citet{Ansorge89}.\label{fig3}}
\end{figure}
 
\section{Simulation of the Diffractive Interaction}

The diffractive interaction was introduced 
to particle physics, in 1960, by 
\citet{GoodWalker60}
as an extension of the theoretical interpretation 
of deuteron dissociation on the nuclear target 
by \citet{FeinbergPomeranchuk56}.  In our case
the incident proton transitions, after colliding with the target proton,
to an excited state with mass 
$M^*$, and dissociates to a nucleon and multiple pions.
The excited state has the same isospin as the proton.  
The mass difference, $M^* - M_p$, is much smaller than the total 
center-of-mass energy, $E_{CM}=\sqrt{s}$.  
The momentum transfer ($q_{||}$) is small and parallel to 
the incident momentum ($p_p$):
$q_{||} = (M^{*2}-M_p^2)/2p_p$.  Almost the full momentum of
the projectile is carried by the excited state.
The projectile or target
proton goes to an excited state with the same probability:  
the former is called the projectile diffraction and the latter
the target diffraction.  The double diffractive interaction 
refers to the case when both transition to excited states:
its cross-section is roughly the same as that of
the projectile diffraction. 

In mid 1960's, both single and double diffraction phenomena have been
observed experimentally \citep{AlberiGoggi81}.
As has been describe earlier, this interaction
has been found to grow in cross-section with the incident 
energy as shown by the curves labeled as ``Diff (all)''
and ``Diff (single)'' in Fig.1a \citep{DiffDissoc1,DiffDissoc2,
DiffDissoc3,DiffDissoc4,DiffDissoc5,DiffDissoc6}.
We note here that the highest energy data point 
in Fig.1a for the single diffraction come from the collider
at CERN where forward-most particles pass through
the beam pipe.  This leaves some uncertainty in obtaining 
the angle-integrated cross-section and $M^{*2}$ distribution
(see, eg., \citet{Donnachie04}).

Our program (see footnote 2) first chooses one from the projectile, 
target and double
diffractions.  In the program,
the single diffraction cross-section is
taken from Fig.1 of \citet{Goulianos95} and the double
from Fig.4 of \citet{CDF01}.  
We impose a suppression factor for the double diffraction
for $T_p<31$ GeV according to the formula given
by \citet{Givernaud79}.  

The $M^{*2}$ distribution of the exited state is simulated as a sum
of 2 resonances (at 1400 and 1688~MeV) and one continuum tail 
proportional to $1/M^{*2}$ \citep{Goulianos83}.
Fig.4a shows the $M^{*2}$ distribution our program produces, 
which approximates that shown in Fig.1 of \citet{DiffDissoc6} or 
Fig.25 of \citet{Goulianos83}.  The sharp
spike at the pion threshold ($M^{*2} \sim 1.2$~GeV) is an artifact
in the program and does not affect the results presented here. 

\begin{figure}
\epsscale{.80}
\plotone{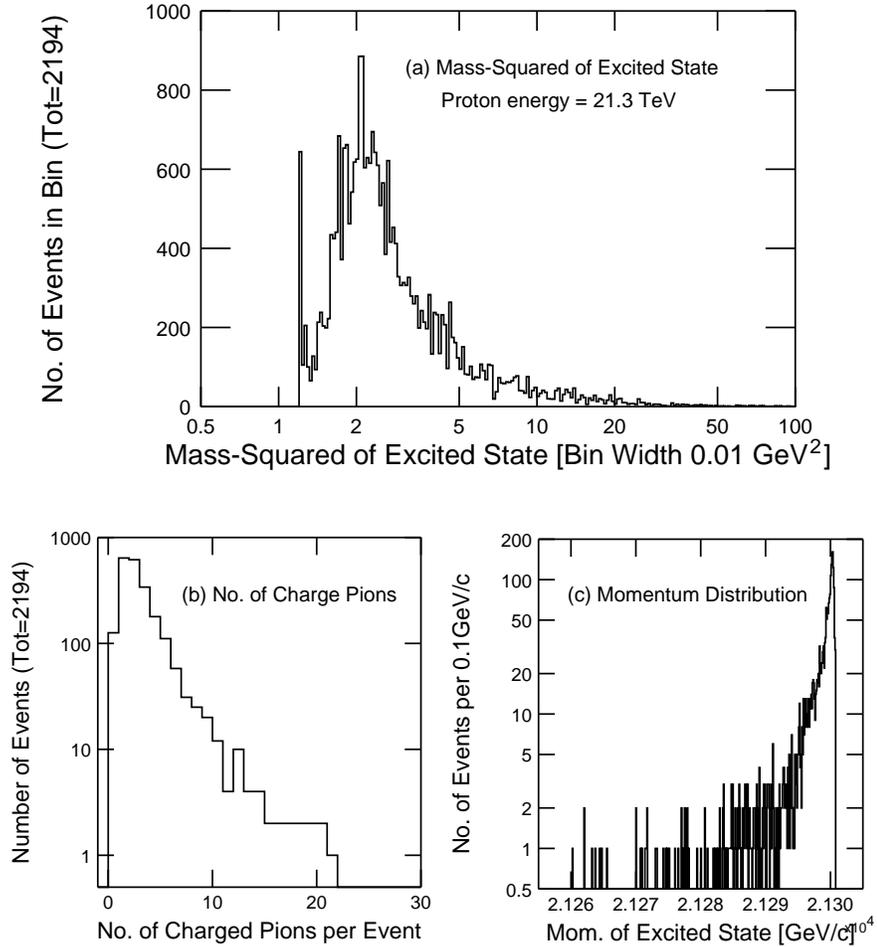}
\caption{Distributions of 3 quantities in projectile 
diffraction events at $T_p=21.3$~TeV 
generated by DiffDissocSimNew.py: 
(a) $M^{*2}$ mass, (b) charged pion multiplicity, and 
(c) laboratory momentum of $M^*$.  Note that a sharp spike 
in (a) at mass-squared 1.2~GeV$^2$ is an artifact.
\label{fig4}}
\end{figure}

For a $M^*$ chosen, a charge multiplicity is selected
from the normal distribution with average of 
$n_0=2.0\sqrt{M^*-M_p}$ (masses in GeV) and $rms$ of $n_0/2$
\citep{Cool82,Goulianos83}.
We then choose a $\pi^0$ multiplicity 
with average of $n_0/2$ and $rms$ of $n_0/4$.  Fig.4b shows
the charged multiplicity distribution
for the single diffraction of $T_p = 21.3$~TeV or 
$E_{CM} = 200$~GeV.  This can be compared with that for 
the non-diffractive process at the same energy shown in Fig.3.

The charged multiplicity is divided between $\pi^+$ and $\pi^-$
under the charge conservation: for an odd charged multiplicity
we make one more $\pi^+$ than $\pi^-$; for an even charged
multiplicity, we make an equal number of $\pi^+$ and $\pi^-$, 
implying the 
associated baryon is proton.  Available energy, 
$M^*-M_p-nM_\pi$, will be divided randomly among the pions.
Note that the kinetic energy going to the nucleon can be neglected
because $M_p \gg M_\pi$ in the CM system of $M^*$. 
 
Finally the momentum of the excited state is calculated 
on the basis of energy-momentum conservation: its distribution 
peaks sharply toward the momentum the projectile (see Fig.4c),
in this case, 21.3~TeV/c in the laboratory frame.
The pion momenta in the CM system of $M^*$ are then transformed
to the laboratory frame, and the neutral pions among them 
are forced to decay to gamma-rays.

Gamma-ray spectra produced by our diffractive interaction simulator
are compared with those by the model A non-diffractive process 
in Fig.2b.  Note that the projectile diffraction 
contributes in a narrow energy band at around one-tenth of $T_p$ 
while the target diffraction piles up 
in a narrow energy band around 100~MeV almost independently of 
the incident proton energy. 

We note that the diffractive dissociation has also been implemented
in Pythia 6.1/6.2 \citep{Pythia62} and in PHOJET 
(see footnote 6).   The particle interaction modeling of our model A
is similar to that used in Pythia:
in fact the total charge multiplicity distribution  
including the double diffraction by Pythia 6.2 
with the higher order terms is similar to that of model A 
shown in Fig.3.  Detailed comparison on the difference between 
the diffractive parts of Pythia and PHOJET is given
in a review by \citet{Guillaud&Sobol04}.  We
note that the diffractive parts of Pythia and PHOJET are optimized
in combination with their respective non-diffractive counter parts.   

\section{Gamma-Ray Spectra for Model Spectra of Galactic Protons}

Before proceeding to calculate the gamma-ray spectra produced
by protons with continuum spectra, we have summed the $\pi^0$ yields
from the non-diffractive and diffractive interactions to 
compare with the experimental data listed in \citet{Dermer86}.
Fig.5 shows the inclusive $\pi^0$ cross-section or, equivalently,
the average multiplicity of $\pi^0$ per $p$-$p$ interaction
multiplied by the cross-sections for $pp(\bar{p}p)$ collisions: 
the solid line,
labeled as model ``A All'', is the sum of the non-diffractive and
double diffractive contributions; the dashed line,
labeled as model B, represents the non-diffractive contribution
without the multiple parton interaction terms.
We note that contribution of the diffractive process,
model A diff, is small
in the multiplicity, just as in the charged multiplicity
shown in Fig.3.  Model A reproduces the measured $\pi^0$ multiplicity
quite well above the resonance region ($T_p>3$ GeV).  
We also note that the scaling model of
\citet{Dermer86} as well as our model B reproduce the data well.

\begin{figure}
\epsscale{.80}
\plotone{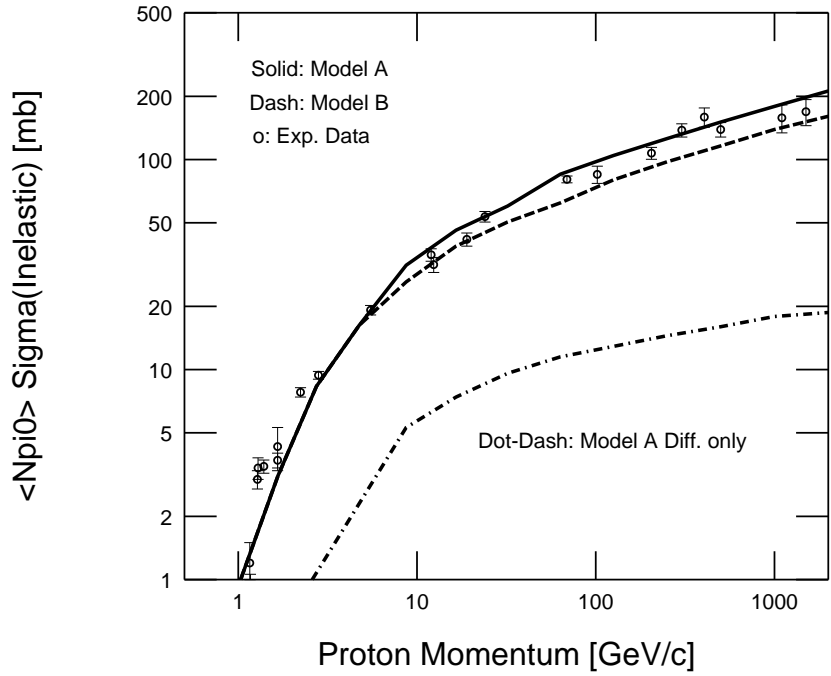}
\caption{Averaged neutral pion multiplicity for the $p$-$p$ 
and $p$-$\bar{p}$ inelastic interaction. Curves are for 
model ``A All'' ({\it{solid}}), 
model B ({\it{dashed}}), and model A diffractive ({\it{dot-dashed}}).  
Data are from Table 1 of \citet{Dermer86}.\label{fig5}} 
\end{figure}

To the accuracy needed for the present study, continuum
spectra of protons can be approximated
by a sum of a series of mono-energetic proton beams.
We choose a geometrical
series of $T_p=1000.0 \times 2^{(N-22)/2}$~GeV where $N=0-40$. 
Each proton energy bin covers from $2^{-0.25}T_p$ to $2^{0.25}T_p$.
The gamma-ray spectra for mono-energetic protons
of kinetic energies $T_p$ 
listed in Column 1 of Table 1 are weighted by the appropriate 
$p$-$p$ cross-sections for the energy given in Columns 2-4.
Figs.2a and 2b show such samples for 3 $T_p$'s for models A and 
B, and for the non-diffractive and diffractive interactions.
The cross-section-weighted gamma-ray spectra for 
mono-energetic protons of $T_p$ is then multiplied with
a proton spectrum factor of the three listed in Column 5-7. 
By summing over all $T_p$'s we obtain the gamma-ray spectrum 
for the proton spectrum.
We note here that the sum over a geometrical series of 
proton kinetic energies ($T_p$) makes the gamma-ray spectrum 
corresponding to a proton spectrum with power-law index 1.0.  
The 3 proton spectral factors in Table 1 are accordingly adjusted 
and mutually normized to 1.0 at $T_p=1$~TeV:
they will be normalized differently, as will be described below, 
when the gamma-ray fluxes are obtained.

Among many model spectra proposed for the Galactic protons,
we have chosen the power-law spectrum of index 2.0, the local interstellar
spectrum, and a trial spectrum for the Galactic ridge region.  
We give short description on them below.  
\begin{description}
 \item[Ind2:] Power-law spectrum with index 2.0 representing the 
       proton spectrum at acceleration sites (eg. \citet{Ellison04}).
 \item[LIS:] Local Interstellar Spectrum (LIS) defined as the
       spectrum just outside of the solar system.  It can be obtained
       by removing the solar modulation effect from cosmic ray 
       spectra measured around the Earth.  Our LIS has been taken
       from the solid curve in Fig.4 of \citet{MSOP02} that has 
       been obtained as a result of fitting GALPROP to various 
       observational data.  
       It takes a power-law spectrum with index 2.7 for $T_p>15$~GeV .
 \item[Trial4GR:] A broken power-law spectrum with indices 2.5 
       ($T_p>20$ GeV) and 2.2 ($T_p<20$ GeV).  Considering
       various uncertainties in the source spectra, propagation,
       and possible reacceleration of protons and electrons, 
       this trial spectrum remains to be a possibility 
       in the Galactic ridge region.
\end{description}
The total gamma-ray spectra for Ind2, LIS and Trial4GR are 
presented after multiplying with $E_\gamma^2$ 
for models A and B in Figs.6a, 6b and 6c.
We note that model A gives harder gamma-ray spectra 
than model B for all 3 proton spectra for $E_\gamma>1$~GeV (Fig.6):
the asymptotic power-law indices for model A/model B are 
$1.96/2.03$ for Ind2, $2.65/2.71$ for LIS and $2.47/2.53$ for Trial4GR.
It is to be noted that the model A gamma-ray spectra
are harder than those of incident protons by index $\sim 0.05$.

\begin{figure}
\epsscale{1.0}
\plotone{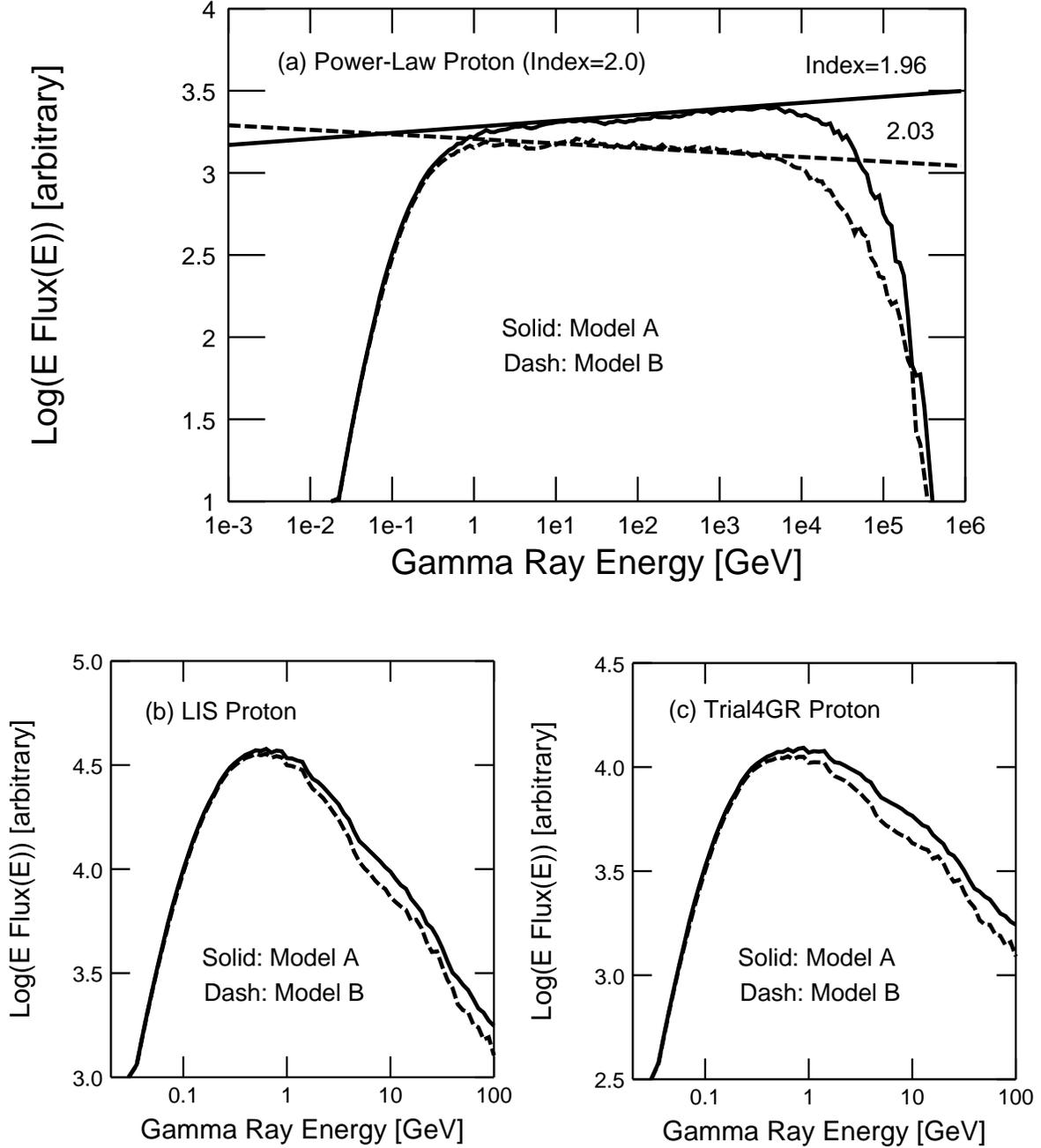}
\caption{Gamma-ray spectra predicted for the 3 proton spectra
between $0.488$~GeV~$<T_p<512$~TeV: 
(a) power-law with index=2.0, (b) LIS, and (c) Trial4GR.
Curves are for model A ({\it{solid}}) and model B ({\it{dashed}}).  
Asymptotic power-law indices of gamma-ray spectra are: 1.96/2.03
(Index=2 model A/model B), 2.65/2.71 (LIS model A/model B),
and 2.47/2.53 (Trial4GR Model A/Model B).
\label{fig6}}
\end{figure}

\section{Comparison with EGRET Galactic Ridge Spectrum}

The EGRET count and exposure maps for the observation period 1-4 
have been downloaded from the EGRET archive (see footnote 3).
All point sources listed in the EGRET 3rd Catalog
\citep{EGRET3rdCatalog} are then removed 
using the point-spread function of EGRET for each energy band 
and for the power-law index of each point source 
listed in the catalog.  The flux between 100~MeV and 10~GeV 
has been constrained to that of each point source
listed in the EGRET 3rd Catalog.
The point-source-subtracted count map is then divided by the
corresponding exposure map to make the intensity map.  
The intensity between the Galactic latitude 
$\pm 6.0$~deg. and Galactic longitude $\pm 30.0$~deg 
has been summed and normalized to a unit solid angle
to become our EGRET Galactic ridge spectrum used in this work.
The point-source-subtracted intensity map 
has been checked to be consistent with the similar map given 
in \citet{SMR00}.  The intensity is then divided by the 
bin width and multiplied by $E_{bin}^2$, or the mid-energy 
squared.  The results are given in the 4th column of Table 3.

A second EGRET spectrum has been calculated in the same manner as above
except that the point-source-subtracted count map is processed
further to deconvolve the point spread function (PSF): 
energy dependency of the PSF used in the deconvolution has been
derived assuming a power-law incident gamma-ray (index 2.1).
Details will be described in a separate publication
(T. Kamae et al. 2005, in preparation).  
The deconvolution removes artifacts introduced
by the broad EGRET PSF, in particular, for $E_\gamma<150$~MeV
and allows us to compare the observed spectrum and model 
predictions directly.
Thus obtained EGRET Galactic ridge spectrum (referred to as 
``deconvolved'') is given in the 5th column of Table 3. 

\begin{table}
\begin{center}
\caption{EGRET Spectra of Galactic Ridge}
\end{center}

\begin{tabular}{rrrrr}
\hline\hline
\multicolumn{3}{c}{EGRET Energy Bins} & 
\multicolumn{2}{c}{$log_{10}(E_\gamma^2 Intensity)$} \\
$E_{min}$ & $E_{max}$ & $Log(E_{center})$ & Intensity Map & Deconv. Int. Map \\
(MeV) & (MeV) & (GeV) & (GeV/cm$^2$/sr/s) & (GeV/cm$^2$/sr/s) \\
\hline 
30 & 50 & -1.411 & -4.7969 & -4.6875\\
50 & 70 & -1.227 & -4.8109 & -4.7406 \\
70 & 100 & -1.077 & -4.7140 & -4.6631 \\
100 & 150 & -0.911 & -4.6005 & -4.5674 \\
150 & 300 & -0.673 & -4.4144 & -4.3987 \\
300 & 500 & -0.411 & -4.3384 & -4.3313 \\
500 & 1000 & -0.150 & -4.2367 & -4.2367 \\
1000 & 2000 & 0.150 & -4.1924 & -4.1924 \\
2000 & 4000 & 0.451 & -4.2538 & -4.2538 \\
4000 & 10000 & 0.801 & -4.4153 & -4.4153 \\
\hline\hline
\end{tabular}

\end{table}

The numbers of gamma-rays in our Galactic ridge region 
before the deconvolution are:
2768, 7700, 18656, 26784, 50081, 27469, 22601, 10971, 4217, and 1235 for the
10 EGRET energy bands in the ascending order in energy.  
The statistical uncertainty (fwhm) is 4\% or
better except for the lowest and highest energy bands.
However, possible systematic error associated with the EGRET spectrum is 
hard to assess: we assume it to be 
$\pm 15$\% as has been done in \citet{SMR04}.

When steep spectra (eg. power-law index $\sim 2.0-3.0$) are
histogrammed in wide energy bins as had been done in the EGRET data
analysis, some systematic bias may be introduced.
We studied this bias by binning model A predictions for 
LIS and Trial4GR into the EGRET bins and multiplying with 
the mid-energy of the bins.  The coarsely-binned data tend to give
higher values of 
$E_\gamma^2 \times Flux(\gamma)$($=E_\gamma \times Flux(E_\gamma)$) 
but the difference 
is only 5-10\% and negligible in the accuracy of the present work.

We proceed to calculate the $E^2_\gamma$-weighted gamma-ray flux
(equivalent to $E_\gamma^2 \times Flux(\gamma)$) for EGRET,
model A with LIS, model B with LIS, the GALPROP result 
obtained with the conventional parameter setting 
where the cosmic-ray proton spectrum becomes LIS
\citep{SMR04}, and the result given in \citet{SB81}.
We note that the GALPROP model described
in \citet{SMR00}, \citet{MSOP02}, and \citet{SMR04} 
with the conventional parameter setting 
give similar gamma-ray spectrum.  We refer to
the one in \citet{SMR04}
because it is produced with the current (C++) version of GALPROP.

As has been stated in Sec.1, the gamma-ray spectrum of
EGRET should be analyzed including the contributions
from bremsstrhlung and inverse-Compton.  However,
the GeV Excess was first noted by comparing the shape
of $E_\gamma^2$-weighted EGRET spectrum with that obtained on the 
$\pi^0$ inclusive cross-section based on the scaling hypothesis.  
Because of this history, we focus on the spectral shape in Fig.7 
and normalize the 4 model $E_\gamma^2$-weighted gamma-ray spectra 
in the energy range $E_\gamma < 300$ MeV where the
$\pi^0 \rightarrow \gamma$ spectrum becomes insensitive to the
incident proton spectrum.  The EGRET data has been normalized
so that the average of the $150-300$~MeV and $300-500$~MeV bins
to agree with the model A prediction. 
We note that the proton spectrum 
assumed in \citet{SB81} has a power-law index of 2.75 and 
a break at $T_p=7$~GeV.  

\begin{figure}
\epsscale{.80}
\plotone{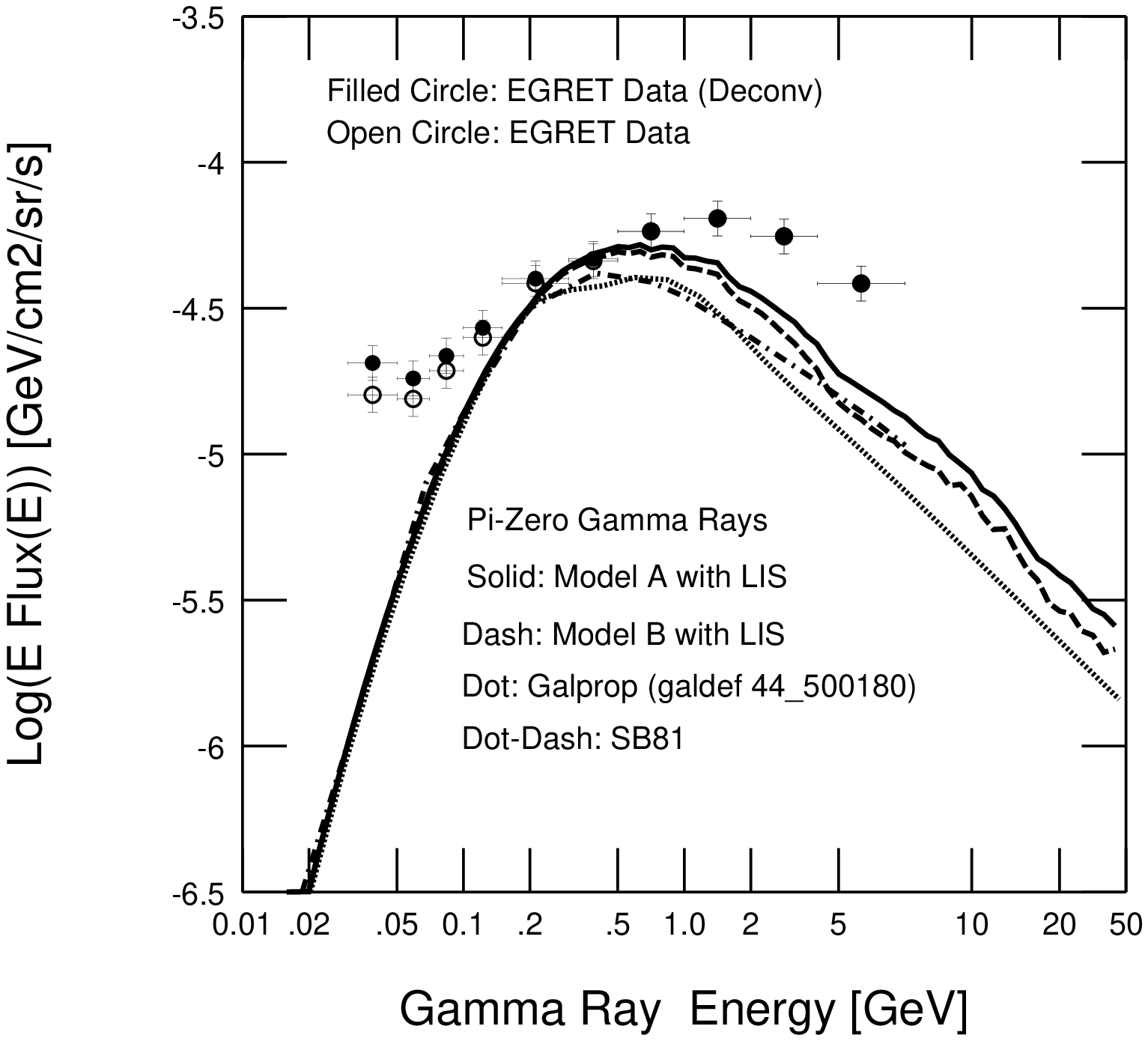}
\caption{Model $\pi^0$ gamma-ray spectra and the EGRET data:
The filled circles are for the PSF deconvolved
EGRET spectrum; and the open circles are for the EGRET spectrum, both 
for the Galactic Ridge ($-6< b <6$~deg, $-30< \ell <30$~deg).  
Model curves are for model A with LIS ({\it{solid}}),
model B with LIS ({\it{dashed}}), \citet{SB81}({\it{dot-dashed}}), 
and GALPROP with the conventional cosmic
ray spectra \citep{SMR04}({\it{dotted}}).  
Model spectra are mutually normalized in $E_\gamma < 300$~MeV.
The EGRET data ({\it{open circle}}) are normalized to the model A with LIS
in $150 < E_\gamma < 500$~MeV.}
\label{fig7}
\end{figure}

Model A with the LIS spectrum 
predicts a broad peak at around $\sim$0.8~GeV for gamma-rays
of $\pi^0$ origin: its spectral shape is closest  
to that of the EGRET ridge data among the four shown in Fig.7.  
The proton spectrum in the
Galactic ridge region is considered to lie between that of
the source speculated to be close to that of power-law with 
index $\sim 2.0$ \citep{Ellison04}
and that observed at $\sim 8.5$kpc away, a power-law spectrum
with index $\sim 2.7$.  
When we compare model B with LIS ({\it{dashed}}) with \citet{SMR04}
({\it{dotted}}) and 
\citet{SB81}({\it{dot-dashed}}), we note that model B 
produces more higher energy gamma-rays than the others.
This is interpreted as due to the fact that Pythia 
without the higher order terms over-produces $\pi^0$ 
in the forward-most phase-space region or the highest energy region
as noted by \citet{Mori97}.  We note that the asymptotic power-law 
index of model B is similar to that of \citet{SMR04}.

As the final step of analysis, we combine the model A
prediction with the bremsstrahlung and inverse-Compton spectra
predicted by GALPROP with the conventional cosmic ray
spectra (the parameter ``galdef 44\_500180'' in \citet{SMR04}).   
Here we normalize the model A $\pi^0 \rightarrow \gamma$-ray 
spectra (with LIS and Trial4GR) to $\pi^0$ gamma-ray spectrum 
of this GALPROP model in the energy region $E_\gamma < 300$~MeV 
as has been done in Fig.7.  
Since the contributions by pion decay, bremsstrahlung, and inverse Compton
are mutually fixed within the GALPROP model, we add
the model A (Trial4GR and LIS), the GALPROP bremsstrahlung, 
and the GALPROP inverse Compton to obtain the spectra labeled as
``model A with Trial4GR'' ({\it{solid curve}}) and ``model A with LIS''
({\it{dashed curve}}).   We note that the normalization to the EGRET data 
relative to the 3 models is still unconstrained
and our focus should be on the spectral shape. 

In Fig.8 we note that the peak energies and the widths of 
the two model A spectra (LIS and Trial4GR) are closer to 
those of the EGRET data than those of the GALPROP with the conventional
cosmic-ray spectrum (galdef 44\_500180):
the discrepancy in the GeV region or ``GeV Excess'' is reduced to 
about 50~\% if we compare model A (LIS) and the GALPROP spectrum.  
When compared with the spectra in Fig.7, the addition of the 
bremsstrahlung contribution
shifts the peaks in $E^2_\gamma F(\gamma)$ to lower energies.  
The EGRET spectrum deconvolved of the 
point spread function improves the agreement between 
the data and the models slightly
in the lower slope of the gamma-ray spectrum.  
The proton spectrum Trial4GR combined with model A 
(Fig.8, {\it{solid curve}}) produces a $E^2_\gamma F(\gamma)$ spectrum
consistent with that of the EGRET data in the GeV range. 

\begin{figure}
\epsscale{.80}
\plotone{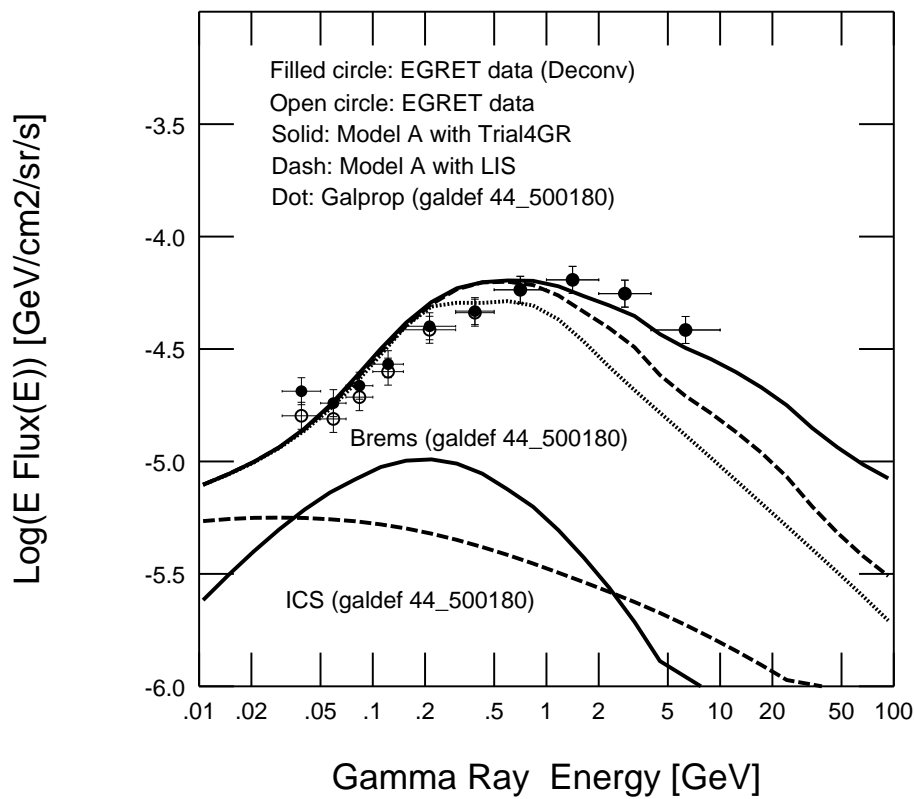}
\caption{Model gamma-ray spectra including the contributions
from bremsstrahlung and inverse-Compton and the EGRET data:
Data labels are same as in Fig.7.  Model curves are
:(Brems) bremsstrahlung contribution; (ICS) inverse-Compton 
contribution, of GALPROP with parameters galdef 44\_500180 
in \citet{SMR04}.  Other curves are: 
model A (Trial4GR)$+$Brems$+$ICS ({\it{solid}}); 
model A (LIS)$+$Brems$+$ICS ({\it{dashed}}); 
$\pi^0+$Brems$+$ICS by GALPROP with galdef 44\_500180 
\citep{SMR04} ({\it{dotted}}).}
\end{figure}

The higher $\pi^0$ yield in model A relative to model B
implies higher anti-proton yield.  
Our study (T. Kamae et al. 2005, in preparation) shows that Pythia 6.2 with 
the higher order terms, with the model A cross-section, and 
with the LIS proton spectrum produces $\sim1.5-2.0$ times more
secondary anti-protons than that without
the higher order terms, with the model B cross-section, and
with the LIS proton spectrum, for $E(\bar{p})=1-20$~GeV (see
Fig.9).  Several comments are in order.  
The non-diffractive inelastic process is expected to produce
$\bar{p}$ even for $T_p<62.5$~GeV.  However our low energy model 
based on \citet{SB81} and \citet{Blattnig00} has not been implemented
with the $pp \rightarrow \bar{p}$ inclusive process.  Hence we have used
Pythia 6.2 with the higher order terms and Pythia 6.1 
to $T_p=10$~GeV 
for models A and B, respectively, to obtain the $\bar{p}$
yield shown in Fig.9.

\begin{figure}
\epsscale{.80}
\plotone{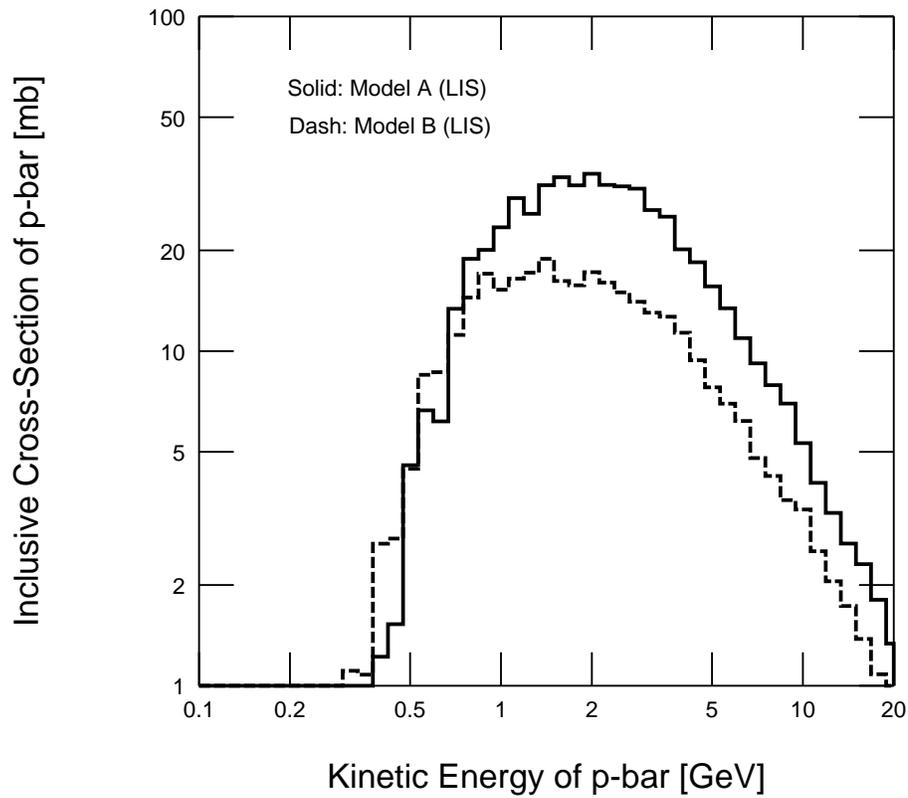}
\caption{Anti-protons spectra predicted for the local cosmic ray proton
spectrum (LIS) by model A ({\it{solid curve}}) and model B 
({\it{dashed}}).  The bin
width is 5~\% of proton kinetic energy.  The low energy non-diffractive
model of \citet{SB81} has been replaced by Pythia for this 
calculation (see text).}
\end{figure}

\section{Conclusion and Future Prospect}

We conclude on the analyses presented here that
an accurate modeling of the $p$-$p$ interaction (model A)
with the diffractive process and the Feynman scaling violation
makes the gamma-ray spectrum harder
and produces 30$-$80\% more gamma-rays (Figs.6, 7, and 8)
than previous predictions \citep{Strong78,SB81,Dermer86,Stecker89,Mori97}
for incident protons with $T_p>100$~GeV.  
Combination of the two can explain $\sim 50$~\% of 
the ``GeV Excess'' in the EGRET Galactic 
ridge spectrum within the conventional cosmic proton and electron
spectra as shown in Fig.8.  The above statement is only relative
to other $pp \rightarrow \pi^0$ production models: the absolute 
prediction of the Galactic ridge gamma-ray spectrum 
is contingent on the absolute normalization, or the absolute 
cosmic ray fluxes, the absolute ISM density, and the absolute radiation
field density.  As far as the gamma-ray spectral shape is concerned,
the remaining discrepancy (50\%) requires some modification to 
the conventional cosmic ray spectra: one possibility is to
assume the proton spectrum in the Galactic ridge to be 
a little harder than that of the solar neighborhood, eg.
$\sim 2.5$ in power-law index as Trail4GR in Fig.8.  

We have compared model A critically with data from
accelerator experiments (Figs.1a, 3, 4, and 5) and  
confirmed that important aspects of experimental data are reproduced 
much better by model A than model B which crudely reflects
the Feynman scaling hypothesis.  We believe that all
future cosmic $p$-$p$ interaction models must include
the diffractive process and incorporate the scaling violation. 

Model A with LIS predicts the $\bar{p}$ flux to be higher
by a factor of $\sim 1.5-2$ relative to model B.  
We will compare the above prediction with 
the recent measurements on the $\bar{p}$ flux
\citep{Orito00, Asaoka02, Basini99, Boezio01}
and the prediction by \citet{MSOP02} in a separate publication
(T. Kamae et al. 2005, in preparation). 

In this work, we have neglected contributions of $\alpha$ particles 
and helium atoms/ions to the gamma-ray spectrum.  
We note that the abundance of helium atoms/ions is about 7\%
of hydgrogen atoms/ions and so is the $\alpha$ to $p$ ratio. 
The spectral index of $\alpha$ particles is known to be
lower by about $0.1$ (see eg., \citet{Mori97,Schlickeiser02}).  
Since there are no measurement
on the $\pi^0$ inclusive cross-section for 
$pHe$, $\alpha p$, and $\alpha He$ interactions at high energies, 
we refer to an estimate on possible deviation from that of $p$-$p$.  
We take the results obtained on $pd$ and $p$-$p$ interactions 
by \citet{DiffDissoc6}.  From the reference,
we learn that: the $(\sigma_T(pd)/\sigma_T(pp))^2$ remains constant
over the incident momentum range of the experiment;
and the coherent factor in the diffractive process is less than 10\%.  
Since the coherent factor decreases rapidly as the momentum transfer
increases, we expect it to be much smaller than 10~\% for 
the non-diffractive process. 
By inference we conclude that the gamma-ray spectra produced 
by $pHe$, $\alpha p$, and $\alpha He$ interactions can safely be 
represented by those by $p$-$p$ interactions as has been done in this work.

\begin{figure}
\epsscale{.80}
\plotone{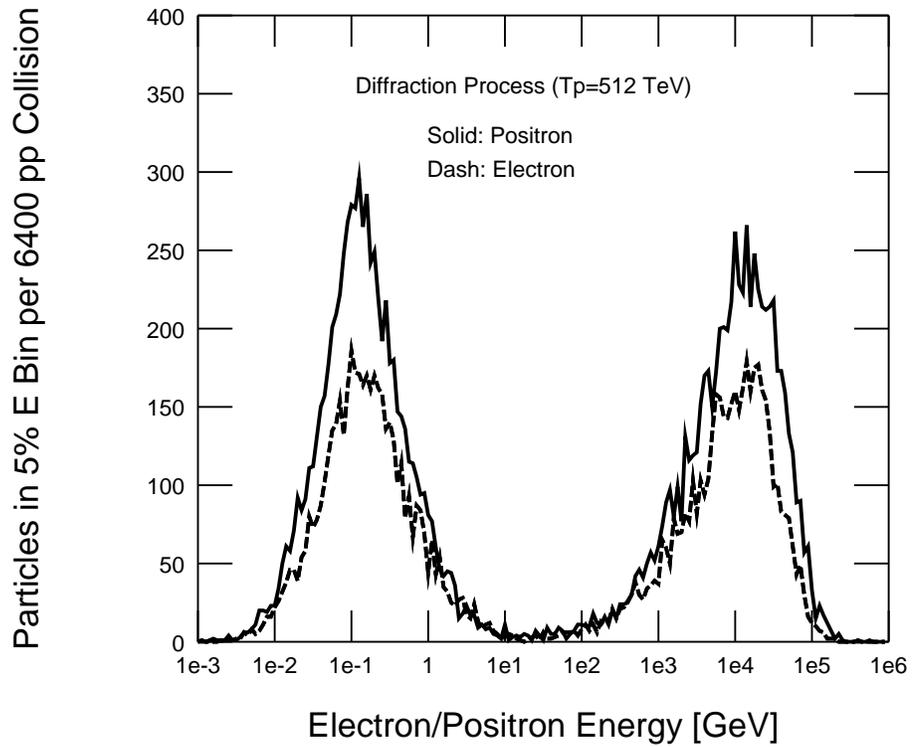}
\caption{Spectra of electrons ({\it{dashed curve}}) and 
positrons ({\it{solid curve}}) 
produced in 6400 diffractive interactions by protons with $T_p=512$~TeV.}
\end{figure}

The findings of this work make the following interesting 
predictions on the spectra of other cosmic ray
particles (T. Kamae et al. 2005, in preparation):
\begin{itemize}
\item The diffractive interaction and scaling violation (model A) 
      nearly double the TeV gamma-ray yield from interaction of
      power-law protons with index=2.0,
      compared to that predicted with model B or our approximation
      to the scaling model (see Fig.6a).
      We also expect all decay products of charged pions ($e^+$, $e^-$, 
      $\nu_e$, $\bar{\nu_e}$, $\nu_\mu$, and $\bar{\nu_\mu}$) 
      to increase by a similar proportion 
      near the highest end of their spectra.
\item Combination of the inherently low multiplicity of the 
      diffractive process and the charge
      conservation predicts $\sim 1.5$ times as many $\pi^+$ as $\pi^-$
      for $T_p=512$~TeV protons as shown in Fig.10.
      We hence expect $e^+/e^-$ and $\nu_e/\bar{\nu_e}$ 
      to increase near the highest end of their spectra
      where the diffractive process contributes most.  
      We note that possible increase of $e^+$ relative to $e^-$ has
      been reported in the cosmic $e^+$ spectrum above 5~GeV. 
      \citep{Coutu99}. 
\end{itemize}

\acknowledgments

One of the authors (Kamae) would like to acknowledge valuable discussions 
with Igor Moskalenko, Andy Strong and Dino Goulianos.  
Moskalenko informed him of the possible excess observed 
in the $e^+$ spectrum \citep{Coutu99}.  
Moskalenko and Strong kindly made the GALPROP
program and galdef parameter files available to him 
and answered many questions about the program.
Goulianos directed him to several references.  The authors acknowledge
the referee for informing them of PHOJET, making valuable suggestions
to improve the description, and correcting errors. 
T. Mizuno, I. Moskalenko and K. Goulianos kindly read an early version 
of the manuscript and helped to make this paper readable both for 
astrophysicists and particle physicists.
The authors gratefully acknowledge advice and assistance given by 
D. Elwe, J. Chiang, S. Digel, I. Grenier, P. Kunz, G. Madejski, H. Tajima
P. Nolan, R. Hartman, S. Hunter, M. Mori, J. Ormes, 
F. Stecker, D. Thompson, M. Asai, N. Graf and Y. Shimizu.  

\appendix

\section{Parameter Settings of Pythia for Models A and B}

Pythia has been used to calculate the $\pi^0 \rightarrow \gamma$ 
inclusive cross-section for $T_p\ge 62.5$~GeV.
The following parameter settings of Pythia have been used for models A
and B.  The gamma-ray spectra generated by model A
are shown in Fig.11 for several monoenergetic proton beams.
\begin{itemize}
\item Model A: Pythia 6.2 with the following parameter setting 
               \citep{Pythia62,CDFtuneA}.
  \begin{itemize}
     \item Setup for the multiple interaction ``CDF tune A'': \\
      MSTP(81) = 1       \\
      PARP(82) = 2.0     \\
      PARP(89) = 1800.0  \\
      PARP(90) = 0.25    \\
      MSTP(82) = 4       \\
      PARP(83) = 0.5     \\
      PARP(84) = 0.4     \\
      PARP(85) = 0.9     \\
      PARP(86) = 0.95    \\
      PARP(67) = 4.0     \\
     \item The default setup for $p$-$p$ interaction by setting MSEL in 
           COMMON/PYSUBS/ to 1: MSEL=1      
     \item Force charged pions, charged kaons, K-long's and muons 
           to decay instantly:\\
           KCPI = PYCOMP(211) \\
           MDCY(KCPI,1) = 1 \\
           KCK = PYCOMP(321) \\
           MDCY(KCK,1) = 1 \\
           KCKL = PYCOMP(130) \\
           MDCY(KCKL,1) = 1 \\
           KCMU = PYCOMP(13) \\
           MDCY(KCMU,1) = 1 \\
  \end{itemize}
\item Model B: Pythia6.1 with the following parameter setting \citep{Pythia62}.
  \begin{itemize}
     \item The default setup for $p$-$p$ interaction by setting MSEL in 
           COMMON/PYSUBS/ to 1: MSEL=1      
     \item Force charged pions, charged kaons, K-long's and muons 
           to decay instantly:\\
           KCPI = PYCOMP(211) \\
           MDCY(KCPI,1) = 1 \\
           KCK = PYCOMP(321) \\
           MDCY(KCK,1) = 1 \\
           KCKL = PYCOMP(130) \\
           MDCY(KCKL,1) = 1 \\
           KCMU = PYCOMP(13) \\
           MDCY(KCMU,1) = 1 \\
  \end{itemize}
\end{itemize}

\begin{figure}
\epsscale{.80}
\plotone{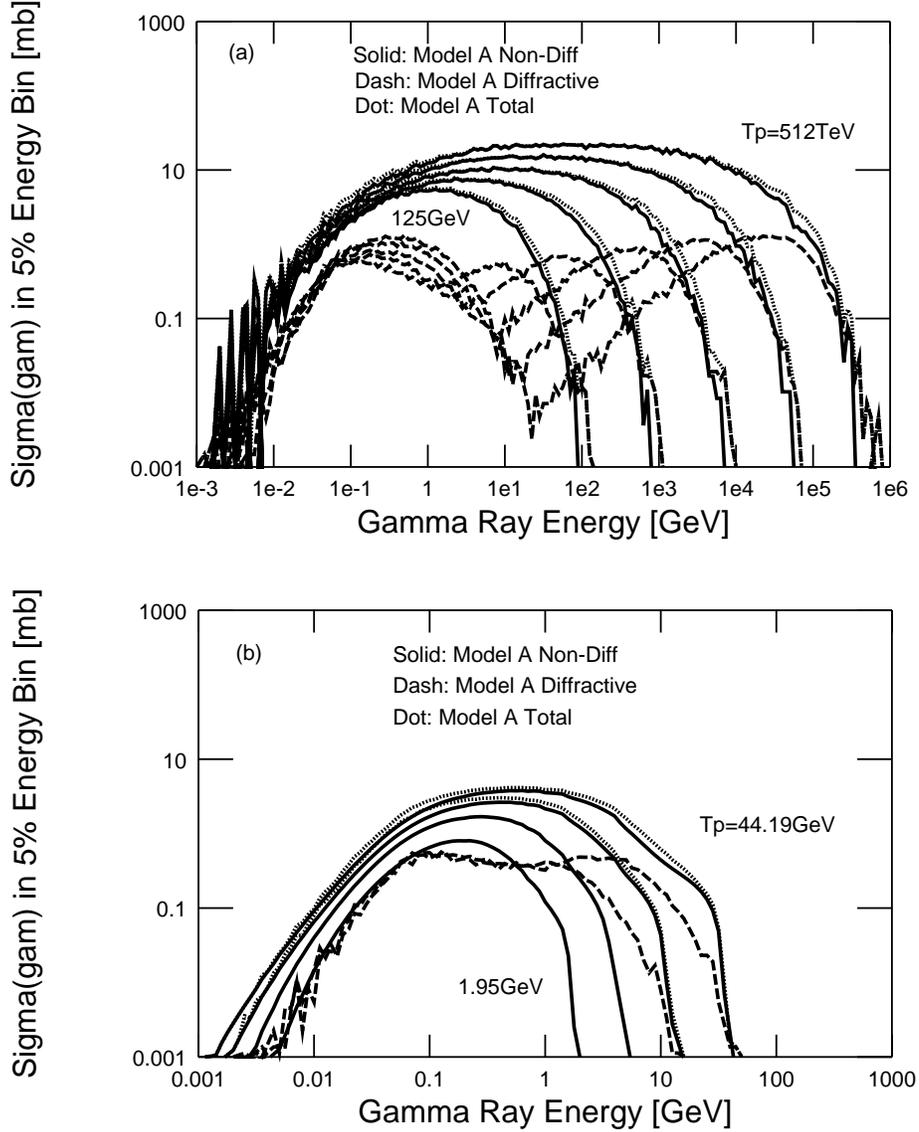}
\caption{Inclusive gamma-ray cross-section per 
$\Delta E_\gamma = 0.05 E_\gamma$ bin
for mono-energetic proton beams: model A non-diffractive ({\it{solid curve}}), 
model A diffractive ({\it{dashed curve}}), 
and model A total ({\it{thin dotted curve}}).  
Proton kinetic energies ($T_p$) are, from right to left:
(a) 512~TeV, 64~TeV, 8~TeV, 1~TeV, and 125~GeV;
(b) 44.19~GeV, 15.63~GeV, 5.52~GeV, and 1.95~GeV.
Note that the diffractive contributions give double humps
at higher energies.  Also note that the cross-section
for the diffractive process is zero for $T_p=$ 5.52~GeV and 1.95~GeV.
\label{figA1}}
\end{figure}

\section{Low Energy Formulae for the Non-Diffractive Process 
for Models A and B}

The cross-sectional formulae of \citet{SB81} parametrized by \citet{Blattnig00}
have been used to calculate the $\pi^0 \rightarrow \gamma$ 
inclusive cross-section for $T_p\le 44.2$~GeV.
\citet{Blattnig00} gives two different parameterizations.  One (No.1 below) 
reproduces the overall $\pi^0$ momentum distribution better but
under-predicts higher energy $\pi^0$ yield in the $p$-$p$ center-of-mass
system.  The other (No.2 below) gives a poorer overall agreement 
and over-predicts higher energy $\pi^0$ yield.
We have mixed the two parameterizations so that the 
$E_\gamma^2$-weighted gamma-ray spectrum connects smoothly to that 
by Pythia at $T_p=62.5$~GeV.

\begin{itemize}
  \item Parameterization No.1 of \citet{SB81}: 
    \begin{itemize} 
       \item $T_p=0.3-2.0$~GeV: Eq. 23 of \citet{Blattnig00}.
       \item $T_p=2.0-44.2$~GeV: Eq. 24 of \citet{Blattnig00}.
    \end{itemize}
 \item Parameterization No.2 of \citet{SB81}:
    \begin{itemize} 
       \item $T_p=0.3-44.2$~GeV: Eq. 32 of \citet{Blattnig00}.
    \end{itemize}   
 \item Mix of Param-1 and Param-2 used in models A and B:
    \begin{itemize}
       \item $\sigma(\pi^0) = (1.0-R)\sigma(\pi^0:$Param-1$) 
                   + R\sigma(\pi^0:$Param-2) \\
            where the ratio $R$ has
            been set to 0.15 to make the gamma-ray 
            spectrum smoothly
            transform from $T_p=44.2$~GeV by the mixed formula to 
            $T_p=62.5$~GeV by Pythia.
    \end{itemize}
\end{itemize}
The gamma-ray spectra generated by the above Mix of Param-1 and Param-2 
are shown in Fig.11 for several monoenergetic proton beams.

\end{document}